\documentclass[aps,prd,preprintnumbers,nofootinbib]{revtex4}
\usepackage[dvipdfmx]{graphicx}
\usepackage{bm,latexsym,amsmath,amsthm,amssymb,amsfonts,color}

\usepackage[normalem]{ulem}

\begin{document} 
\title{\uline{}Area bound for a surface in a strong gravity region}

\author{${}^{1,2}$Tetsuya Shiromizu, ${}^{3}$Yoshimune Tomikawa, ${}^{2,1}$Keisuke Izumi and ${}^4$Hirotaka Yoshino}
\affiliation{${}^1$Department of Mathematics, Nagoya University, Nagoya 464-8602, Japan}
\affiliation{${}^2$Kobayashi-Maskawa Institute, Nagoya University, Nagoya 464-8602, Japan}
\affiliation{${}^3$School of Informatics and Sciences, Nagoya University, Nagoya 464-8601, Japan}
\affiliation{${}^4$Department of Mathematics and Physics, Osaka City University, Osaka 558-8585, Japan}


\begin{abstract}%
For asymptotically flat spacetimes, using the inverse mean curvature flow, 
we show that any compact $2$-surface, $S_0$, whose mean curvature and 
its derivative for outward direction are positive in spacelike hypersurface 
with non-negative Ricci scalar satisfies the inequality $A_0 \leq 4\pi (3Gm)^2$, 
where $A_0$ is the area of $S_0$ and $m$ is the total mass. The upper bound
is realized when $S_0$ is the  photon sphere in a hypersurface isometric to 
$t=$const. slice of the Schwarzschild spacetime.
\end{abstract}

\maketitle

\section{Introduction} 

A black hole is defined as a region that distant observers cannot see. 
It is believed that its typical size is smaller than the Schwarzschild 
radius with the same mass, because only strong gravity in such a compact 
region is expected to have the property of a black hole. One of the 
quantitative arguments is the Penrose inequality \cite{penrose1973}, 
which conjectures that the area of an apparent horizon, $A_{\rm AH}$, 
must satisfy $A_{\rm AH}\le 4\pi (2Gm)^2$ where $G$ is Newton's gravitational 
constant and $m$ is the Arnowitt-Deser-Misner (ADM) mass (in the unit $c=1$). 
The proofs for the Penrose inequality have been given for time-symmetric 
initial data in Refs. \cite{wald1977,bray}.

On the other hand, the existence of an unstable circular photon orbit 
at $r=3Gm$ in the Schwarzschild spacetime is another typical feature 
of strong gravity region. The photon sphere plays a key role for 
gravitational lensing \cite{Virbhadra:1999} or ringdown of waves around 
a black hole \cite{Cardoso:2016}. The photon sphere is located outside 
of the black hole and a photon can escape from such a region in general.
But, since the circular orbit of a photon is also realized by strong gravity 
in a sufficiently compact region, one may expect that an inequality 
analogous to the Penrose inequality should exist also for a surface in a 
strong gravity region inside of a photon sphere.

In this paper, we introduce a new geometrical concept, 
``{\it loosely trapped surface}," as a characterization of a two-dimensional 
surface that exists in a strong gravity region. Then, we prove the inequality 
for its area $A_0$, that is, $A_0 \leq 4\pi (3Gm)^2$. As we see later, 
for the Schwarzschild spacetime, loosely trapped surfaces correspond to 
2-spheres inside of the photon sphere and the photon sphere marginally 
satisfies the definition of a loosely trapped surface. 

The rest of this paper is organized as follows. In Sect. 2, we will 
examine the Schwarzschild spacetime, and the observation there will 
give us a motivation for defining the loosely trapped surface. 
In Sect. 3, we will introduce the notion of the loosely trapped 
surface and present its geometrical property as a proposition which 
will be used later. In Sect. 4, we will prove an inequality analogous 
to the Penrose one for the loosely trapped surface. Finally, we will 
give a summary and discussions in Sect. 5. 

%
%
\section{Lesson from Schwarzschild}

Let us consider the Schwarzschild black hole
with mass $m$. 
The metric is 
\begin{eqnarray}
ds^2=-\Bigl(1-\frac{2Gm}{r}\Bigr)dt^2+\Bigl(1-\frac{2Gm}{r}\Bigr)^{-1}dr^2+r^2 d\Omega_2^2.
\end{eqnarray}
Let us consider a spacelike hypersurface $t=$const..  
The trace of the extrinsic curvature of a $r=$const. sphere
on this hypersurface is given by 
\begin{eqnarray}
k=\frac{2}{r}\Bigl(1-\frac{2Gm}{r}\Bigr)^{1/2}.
\end{eqnarray}
In static or time symmetric slices, this quantity $k$ is proportional to the expansion 
of a null geodesic congruence emitted from the $r=$const. surface. $k$ is zero on the horizon 
$r=2Gm$ and at infinity($r=\infty$). Since the derivative of $k$ with respect 
to $r$ is calculated as 
\begin{eqnarray}
\frac{dk}{dr}=-\frac{2}{r^2}\frac{1-\frac{3Gm}{r}}{\Bigl(1-\frac{2Gm}{r}\Bigr)^{1/2}}, 
\end{eqnarray}
we see that $k$ has the maximum value at $r=3Gm$. This coincides
with the locus of the unstable circular photon orbit, i.e. 
the photon sphere. This property motivates us to introduce new 
concept of a geometrical object as follows. 

On the event horizon $r=2Gm$, everything is trapped and the gravitational 
field is ``infinitely" strong. A strong gravity region is even for outside of 
the event horizon. If photons are emitted isotropically from a point in the 
region between the horizon and the photon sphere, $2Gm<r<3Gm$, more than 50\% 
of the photons will be absorbed to the horizon \cite{synge}. In this sense, photons in this 
region are loosely trapped, and the photon surface is a typical object 
indicating the existence of a strong gravity region. From the above calculation, 
the trace of the extrinsic curvature $k$ is a good indicator for describing 
the strong gravity region in terms of geometrical quantities. To see this, 
let us take two 2-surfaces having the same small positive value $k$. One of them 
is close to the horizon and the other is near spatial infinity. The former 
is in the strong gravity region. To distinguish them, the spatial derivative of 
$k$, $k'(:=dk/dr)$, is useful. More specifically, $k'$ is positive for 
$2Gm < r <3Gm$ (strong gravity region) while it is negative for $r>3Gm$ 
(weak gravity region). Therefore, the positivity of $k'$ indicates the existence 
of the strong gravity. Bearing this observation in mind, we will introduce 
the loosely trapped surface in the next section.

%
%
\section{Loosely trapped surface}

Let us define the {\it loosely trapped surface} as follows:\\
\noindent
{\it Definition:} 
The loosely trapped surface, $S_0$, is defined as a compact 
2-surface such that its trace of the extrinsic curvature $k$ (for the outward 
spatial direction) is positive ($k|_{S_0}=:k_0>0$) 
and the derivative along the outward spatial direction is non-negative ($k'|_{S_0}\ge 0$) in 
a spacelike hypersurface $\Sigma$. 
This definition can be applied for arbitrary initial data 
with the induced metric and extrinsic curvature $(q_{ab}, K_{ab})$ 
satisfying the Hamiltonian and momentum constraints. 
Then, unlike the Schwarzschild spacetime, a loosely trapped surface 
is not necessarily located outside of an apparent horizon in general. 
But, it is still a useful indicator for a strong gravity region.

In this paper, we will suppose the existence of a loosely trapped surface 
although the condition for its existence is rather non-tirivial. As for the 
apparent horizon, the existence has been discussed when matters are very 
concentrated to a small region \cite{schoen}. Similar argument will be interesting. We also 
note that the above definition has the gauge dependence, that is, the 
slice dependence. And, since our definition relies on the observation of 
the Schwarzschild spacetime in the static slice, one may be able to have 
another elegant definition for a strong gravity region. Nevertheless, 
our theorem presented in the next section tells us that there is the 
upper bound for the area of loosely trapped surfaces and then this 
indicates us that our definition works well at least on a spacelike hypersurface 
with non-negative Ricci scalar. We would emphasize that 
our theorem holds regardless of the issue for the naming of loosely 
trapped surface. 

In what follows, we consider the spacelike hypersurface with non-negative 
three-dimensional Ricci scalar, ${}^{(3)}R\ge 0$. We note that on the 
maximal slice where the trace of the extrinsic curvature vanishes, $K=0$, 
the Hamiltonian constraint tells us that this condition is satisfied if  
the energy density is non-negative;
\begin{eqnarray}
  {}^{(3)}R=16\pi G T_{ab}n^a n^b+\tilde K_{ab}\tilde K^{ab} \geq 0,
  \label{positivity-3RiemannScalar}
\end{eqnarray}
where $n^a$ is the future directed unit normal timelike vector to 
$\Sigma$ and $\tilde K_{ab}$ is the traceless part of the extrinsic curvature 
of $\Sigma$. However taking account of the other possibilities with $K\neq 0$
but ${}^{(3)}R\ge 0$, we just require the non-negativity of ${}^{(3)}R$. 

Now, we show that the loosely trapped surface has the following property: \\
\noindent
{\it Proposition:}
The loosely trapped surface in a spacelike hypersurface $\Sigma$ with non-negative Ricci scalar 
has the topology $S^2$ and satisfies $\int_{S_0}k^2dA \leq 16\pi/3$. 
\begin{proof}
Let us consider the outward derivative of the trace of the extrinsic curvature 
$k$ in $\Sigma$, 
\begin{eqnarray}
r^a D_a k  =  -\varphi^{-1}{\cal D}^2 \varphi -\frac{1}{2}{}^{(3)}R+\frac{1}{2}{}^{(2)}R -\frac{1}{2}(k^2+k_{ab}k^{ab}),
\label{r-derivative-meancurvature}
\end{eqnarray}
where $r^a$ is the unit normal spacelike vector to $S_0$, $D_a$ is the covariant derivative of 
$\Sigma$, ${}^{(2)}R$ is the Ricci scalar of $S_0$ and $k_{ab}$ is the 
extrinsic curvature of $S_0$ in $\Sigma$. $\varphi$ is the lapse function for the 
normal direction to $S_0$ in $\Sigma$ and ${\cal D}_a$ is the covariant derivative for $S_0$.

After a little rearrangement of the above and taking the surface integration over $S_0$, 
we have 
\begin{eqnarray}
\frac{1}{2} \int_{S_0} {}^{(2)}R dA & = & \int_{S_0} 
\Bigl[r^aD_ak+\varphi^{-2}({\cal D} \varphi)^2+\frac{1}{2}{}^{(3)}R
+\frac{1}{2}\tilde k_{ab} \tilde k^{ab} +\frac{3}{4}k^2 \Bigr]dA \nonumber \\
& \geq & \frac{3}{4}\int_{S_0} k^2dA >0, \label{2rds}
\end{eqnarray}
where $\tilde k_{ab}$ is the traceless part of $k_{ab}$. Thus, the Gauss-Bonnet theorem 
tells us that the topology of $S_0$ is restricted to $S^2$ and then 
\begin{eqnarray}
\int_{S_0} {}^{(2)}R dA =8\pi. 
\end{eqnarray}
Finally, the inequality \eqref{2rds} implies 
\begin{eqnarray}
\int_{S_0}k^2dA \leq \frac{16\pi}{3}. \label{k-ineq} 
\end{eqnarray}
\end{proof}

Since we did not use the positivity of $k$ on $S_0$ except for the last inequality 
of \eqref{2rds}, we obtain the following Lemma:
{\it Lemma:}Compact 2-surface $\tilde S_0$ with $k'|_{\tilde S_0}\geq 0$  
has the topology $S^2$ and satisfies $\int_{\tilde{S}_0}k^2dA \leq 16\pi/3$ 
if a spacelike hypersurface $\Sigma$ has 
non-negative Ricci scalar ${}^{(3)}R$ and if  
one of
$k'|_{\tilde S_0}$, $k_0$, $\tilde k_{ab}$, ${}^{(3)}R$ and ${\cal D}_a \varphi$ 
is non-zero at least at one point on $\tilde S_0$. 

Moreover, even if we use a relaxed condition for $k'$, 
i.e., $\int_{S_0} r^aD_ak dA \ge 0$ on $S_0$, 
the first inequality in the last line of \eqref{2rds} holds true. 
Namely, the inequality \eqref{k-ineq} 
holds also for the ``averaged loosely trapped surface'' 
satisfying the condition $\int_{S_0}r^aD_ak dA \ge 0$.

%
%
\section{Penrose-like inequality for loosely trapped surface}

We present our main theorem in this section. \\
\noindent
{\it Theorem:}Let $\Sigma$ be an asymptotically flat
spacelike hypersurface with non-negative Ricci scalar ${}^{(3)}R$ and assume 
that $\Sigma$ is foliated by the inverse mean curvature flow 
$\lbrace S_y \rbrace_{y \in {\bf R}}$ ($S_y \approx S^2$). 
Then, the area $A_0$ of a loosely trapped surface $S_0$ in $\Sigma$ 
satisfies the inequality 
\begin{eqnarray}
A_0 \leq 4\pi (3Gm)^2, \label{pi}
\end{eqnarray}
where $m$ is the ADM mass. When the equality holds, 
the region $\Omega$ outside of $S_0$ in $\Sigma$ is isometric to 
the region of $\Omega_{\rm sch}$ outside of the photon sphere in 
the $t=$const. slice of the Schwarzschild spacetime.
\begin{proof}
Let us consider Geroch's energy \cite{geroch1973, wald1977}
\begin{eqnarray}
E(y):=\frac{A^{1/2}(y)}{64\pi^{3/2}G}\int_{S_y}(2{}^{(2)}R-k^2)dA, 
\end{eqnarray}
where $A(y)$ is the area of $S_y$. We consider the evolution of 
$E(y)$ along the inverse mean curvature flow, which is generated by 
the relation $k \varphi=1$ with the lapse function $\varphi$ 
that relates $r^a$ and the coordinate $y$ as 
$r^a=\varphi D^a y$. The first derivative of $E(y)$ is computed as 
\begin{eqnarray}
\frac{dE(y)}{dy} & = & \frac{A^{1/2}(y)}{64\pi^{3/2}G}
\int_{S_y}\Bigl[ 2 \varphi^{-2}({\cal D} \varphi)^2 +{}^{(3)}R+\tilde k_{ab}\tilde k^{ab} \Bigr] dA. 
\label{derivative-Geroch}
\end{eqnarray}
Since we are adopting the inverse mean curvature flow for the foliation, 
$k$ is supposed to be positive definite for each of $S_y$. Then, we have 
\begin{eqnarray}
\frac{dE(y)}{dy} \geq 0 \label{monotonic}
\end{eqnarray}
because of ${}^{(3)}R \geq 0$. Now we take 
the integration of the above between the loosely trapped surface 
$S_0$ and $S_\infty$(i.e., over $\Omega$). 
Because Geroch's energy is reduced to the ADM mass at infinity, we have 
\begin{eqnarray}
m & \geq & \frac{1-\frac{1}{16\pi}\int_{S_0}k^2dA}{4\pi^{1/2}G}A_0^{1/2} \nonumber \\
  & \geq & \frac{A_0^{1/2}}{3 \cdot 2\pi^{1/2} G}. 
\end{eqnarray}
For the last inequality, we used the inequality (\ref{k-ineq}). 
Thus, we have shown the inequality 
\begin{eqnarray}
  A_0 \leq 4 \pi (3Gm)^2.
  \label{Penrose-like-inequality}
\end{eqnarray}

The equality of \eqref{Penrose-like-inequality} occurs 
when ${\cal D}_a \varphi=\tilde k_{ab}={}^{(3)}R=0$ holds on all of 
$S_y$ in $\Omega$ and $r^a D_a k=0$ holds on $S_0$. The condition 
$\mathcal{D}_a \varphi=0$ means the constancy of $\varphi$ on 
each of $S_y$. Due to the inverse mean curvature flow condition $k\varphi=1$, 
the value of $k$ is also constant on $S_y$. 
Therefore, $\varphi$ and $k$ depend 
only on $y$, and Eq.~\eqref{r-derivative-meancurvature} is rewritten as 
\begin{equation}
{}^{(2)}R = 
2k \frac{dk}{dy}+\frac{3}{2}k^2.
\end{equation}
The value of ${}^{(2)}R$ is a constant on $S_y$, and hence, 
each of $S_y$ is spherically symmetric. This implies that 
the region $\Omega$ is spherically symmetric 
and then $\Omega$ is conformally flat. Since 
the induced metric of $\Omega$, $q_{ab}$, is written as $q_{ab}=\psi^4\delta_{ab}$, 
where $\delta_{ab}$ is the metric of flat space, 
the Hamiltonian constraint is reduced to the Laplace equation in the flat space, 
$\partial^2\psi=0$. Therefore, $\psi$ is uniquely determined once the ADM mass $m$
is given, and $\Omega$ is isometric to the region $\Omega_{\rm sch}$ outside of the 
photon sphere in the $t=$const. slice of the Schwarzschild spacetime \footnote{
For a maximally sliced hypersurface, 
${}^{(3)}R=0$ implies $T_{ab}n^an^b=0$ from 
Eq.~\eqref{positivity-3RiemannScalar}. 
If we additionally require the dominant energy condition, 
the whole components of $T_{ab}$ become zero, 
and hence, the spacetime is vacuum in the region $\Omega$. 
Applying Birkhoff's theorem, the time development of $\Omega$ is 
the part $\Omega_{\rm sch} \times \lbrace t \rbrace$ of the Schwarzschild spacetime, 
assuming the absence of outcoming energy flux from inside of the photon sphere.}.
\end{proof}

The formula \eqref{Penrose-like-inequality} corresponds to the Penrose-like 
inequality for loosely trapped surfaces. This upper bound of the area is same 
as that of the photon sphere in the Schwarzschild spacetime.

The Geroch energy was used to show the positivity of the ADM mass \cite{geroch1973}
and the Riemannian Penrose inequality \cite{wald1977} as well. There, 
the monotonicity of $E(y)$ along the inverse mean curvature flow was the 
key ingredient. If one takes the inner boundary to be the central point and 
the apparent horizon, the outcome is the positivity of the ADM mass and the 
Penrose inequality, respectively. Our above proof is a variant of these two works:
the fact that Geroch's quasilocal energy on a 2-sphere $S_y$ is smaller than 
the ADM mass was translated to the bound on the area of $S_y$ using the properties 
of the loosely trapped surface.

Strictly speaking, the foliation by the inverse mean curvature flow has singularities 
in general, and the above proof implicitly assumed the existence of regular inverse 
mean curvature flow. However, one can treat/improve such sensitive part by using tools 
developed in geometric analysis and geometric measure theory \cite{imcf}.

%
%
\section{Summary and discussion}

In this paper, inspired by an unstable circular orbit of a photon 
(i.e., a photon sphere) in the Schwarzschild spacetime, we introduced 
the notion of a loosely trapped surface (Sect. 3)  and proved 
the upper bound of its area that is analogous to 
the Penrose inequality for an apparent horizon (Sect. 4). The upper 
bound is exactly equal to the area of the photon sphere.

Since we employed the inverse mean curvature flow in our proof, 
one may ask whether another method can work for the current issue. 
In the case of the Penrose inequality, after the argument 
based on the inverse mean curvature flow was made \cite{wald1977}, 
Bray's conformal flow \cite{bray} has given a new proof. 
We expect that this would be the case also for the area bound 
for loosely trapped surfaces.

Although the loosely trapped surface is introduced motivated by 
the photon sphere, there are significant differences between these 
two notions. The photon sphere (or its generalization, the photon 
surface \cite{Claudel:2000}) is characterized by the property that 
any photon initially moving tangent to the surface continues to 
move on the same surface. On the other hand, the mean curvature 
$k$ is a purely geometric quantity in a general setup. 
Although $k$ becomes the expansion of null geodesic congruence 
for momentarily static initial data, the photons in this case 
are emitted in the perpendicular directions to the surface. 
However, the fact that the surface with vanishing $r^a D_a k$ 
corresponds to the photon sphere in the Schwarzschild spacetime 
might not be a mere coincidence: we could expect a possible connection 
between the loosely trapped surface and the photon surface 
at a deeper level. It is interesting to explore such a possible connection. 

The relation of the current inequality to the cosmic censorship 
conjecture is also interesting. Since we assumed the regularity 
outside of the loosely trapped surface, finding the violation 
of the inequality indicates the presence of naked singularities. 
Related to this, it is also interesting to ask if the presence of the 
loosely trapped surface implies an occurrence of singularity in the future 
with some global and energy conditions like singularity theorems.

\begin{acknowledgments}
T. S. is supported by Grant-Aid for Scientific Research from Ministry of Education, 
Science, Sports and Culture of Japan (No. 16K05344). 
The work of H. Y. was in part supported by the Grant-in-Aid for
Scientific Research (A) (No. 26247042)
from Japan Society for the Promotion of Science (JSPS).
\end{acknowledgments}


\begin{thebibliography}{9}




\bibitem{penrose1973} 
  R.~Penrose,
  Annals N.\ Y.\ Acad.\ Sci.\  {\bf 224}, 125 (1973).

\bibitem{wald1977} 
P. S. Jang and R. M. Wald, J. Math. Phys. {\bf 18}, 41 (1977). 

\bibitem{bray} 
  H. Bray, J. Diff. Geom. {\bf 59}, 177 (2001).

  
\bibitem{Virbhadra:1999} 
  K.~S.~Virbhadra and G.~F.~R.~Ellis,
  Phys.\ Rev.\ D {\bf 62}, 084003 (2000).

  
\bibitem{Cardoso:2016} 
  V.~Cardoso, E.~Franzin and P.~Pani,
  Phys.\ Rev.\ Lett.\  {\bf 116}, 171101 (2016)
  Erratum: [Phys.\ Rev.\ Lett.\  {\bf 117}, 089902 (2016)].

\bibitem{synge}
For example, J. L. Snyge, Mon. Not. Roy. Soc. {\bf 131}, 463 (1966). 
 
\bibitem{schoen}
R. Schoen and S. T. Yau, Commun. Math. Phys. {\bf 90}, 575 (1983). 

\bibitem{geroch1973} 
R. Geroch, Ann. N.Y. Acad. Sci. {\bf 224}, 108 (1973). 

\bibitem{imcf} 
G. Huisken and T. Ilmanen, J. Diff. Geom. {\bf 59}, 353 (2001). 

\bibitem{Claudel:2000} 
  C.~M.~Claudel, K.~S.~Virbhadra and G.~F.~R.~Ellis,
  J.\ Math.\ Phys.\  {\bf 42}, 818 (2001).

\end{thebibliography}
\end{document}